\documentclass[prl,notitlepage, twocolumn,
showpacs,floatfix,nofootinbib,preprintnumbers,superscriptaddress,amsmath,amssymb]{revtex4-1}

\usepackage{graphicx}\usepackage{epsfig}
\usepackage{amsmath,amssymb}
\usepackage{bm}
\usepackage{color}
\usepackage{float}
\usepackage{dcolumn} 
\usepackage{enumerate}
\usepackage{multirow}
\usepackage{threeparttable}

\begin{document}

\newcommand{\ket}[1]{\left| #1 \right\rangle}
\newcommand{\bra}[1]{\left\langle #1 \right|}
\newcommand{\Fhat}{\hat{F}}

\title{Pairing Nambu-Goldstone Modes within Nuclear Density Functional Theory}

\author{Nobuo Hinohara}
%\email{hinohara@ccs.tsukuba.ac.jp}
\affiliation{%
Center for Computational Sciences, University of Tsukuba, Tsukuba, 305-8577, Japan
}
\affiliation{%
FRIB Laboratory, Michigan State University, East Lansing, Michigan 48824, USA
}
\author{Witold Nazarewicz}
\affiliation{%
Department of Physics and Astronomy and FRIB Laboratory, Michigan State University, East Lansing, Michigan 48824, USA
}

\affiliation{%
  Institute of Theoretical Physics, Faculty of Physics, University of Warsaw,
  02-093 Warsaw, Poland
}

\begin{abstract}
  We show that the Nambu-Goldstone formalism of the broken gauge symmetry in the presence of the $T=1$ pairing condensate offers a  quantitative description of the binding-energy differences of open-shell superfluid nuclei. We conclude that the pairing-rotational moments of inertia are excellent  pairing indicators, which are free from ambiguities attributed to  odd-mass systems.
  We offer  a new, unified  interpretation of the binding-energy differences traditionally 
viewed in the shell model picture as signatures of the valence nucleon properties. We present the first systematic analysis of the off-diagonal pairing-rotational moments of inertia and demonstrate  the  mixing of the neutron and proton pairing-rotational modes in the ground states of  even-even nuclei. Finally, we discuss the importance of mass measurements of  neutron-rich nuclei for constraining the pairing energy density functional.
  \end{abstract}


\maketitle
{\it Introduction}.---Spontaneous symmetry breaking explains the collective properties of atomic nuclei  and provides a straightforward physical interpretation  of experimental observables associated with collective modes.  In atomic nuclei, the Nambu-Goldstone (NG) mode \cite{PhysRev.112.1900,PhysRev.117.648,INC_19_154} connects two frames of reference:
the intrinsic frame, where the symmetry is  broken and the NG mode appears as a zero-energy excitation mode,
and the laboratory frame, where the symmetry is strictly conserved.
The excitation of the NG mode can be observed in the laboratory system as a sequence of quantum states originating from a single symmetry-broken intrinsic state. Incorporating correlations related to the symmetry breaking is essential for many-body theories; see, e.g., the discussion in Ref.~\cite{bes-kurchan}.
One of the typical examples of spontaneous symmetry breaking in atomic nuclei is the nuclear deformation due to the rotational symmetry breaking, as a consequence of the attractive particle-hole correlations \cite{Bohr-MottelsonV2,PGOtten,Nazarewicz94,Frauendorf01}. 
Rotational bands can be viewed  as  NG mode excitations. 

Nucleonic pairing is another common phenomenon in atomic nuclei associated with spontaneous symmetry breaking. Ground states of most nuclei can be well described as  pair condensates, in which the particle number symmetry is broken. Superconducting nuclear states result in a NG mode called the pairing rotation, which is seen experimentally through ground-state sequences  of even-even nuclei \cite{bb66,Bayman69,bb77,Brink-Broglia, Broglia20001}. The topic of pairing rotations 
continues to generate much excitement, especially in the context of neutron-rich nuclei
\cite{Oertzen01,PhysRevC.84.044317,PhysRevLett.96.032501,PhysRevLett.107.092501,PhysRevC.87.054321}.

%EDF
Nuclear density functional theory (DFT) is currently the only available microscopic many-body theory that is applicable to the whole nuclear chart. One of the reasons for its success is the flexibility of the formalism to  naturally incorporate the spontaneous symmetry-breaking mechanism.
The form of the nuclear energy density functional (EDF) is  constrained by symmetry considerations; popular Skyrme EDFs 
are built from  density-bilinear terms 
in both the particle-hole and pairing channels \cite{PhysRevC.69.014316,Roh10}.
Considerably less is known about the pairing EDF, primarily because  of the lack of the  experimental observables that can inform us about the detailed structure  of the pairing EDF.

The order parameter for the superfluid phase is the expectation value of the pair creation operator
that can be related to the observed pair transfer cross section \cite{bb66,Brink-Broglia,Bjerregaard19681}.  
However, the coupling constants in the pairing EDF
are conventionally
fitted so that the
theoretical pairing gaps in even-even nuclei reproduce the experimental odd-even mass differences. Such a strategy has been adopted in recent optimization work \cite{PhysRevC.82.024313,PhysRevC.85.024304,PhysRevC.89.054314},
although the relationship between the pairing gap and the experimental odd-even mass difference is indirect.
Moreover, there exist multiple definitions of  theoretical pairing gaps and there are various prescriptions for extracting the  odd-even mass difference from experiment \cite{Sat98,Dug01,PhysRevC.79.034306}. 
 To avoid  ambiguities, it would be best to calculate the odd-even mass difference directly from the theory. Unfortunately, this involves additional uncertainties pertaining to the definition of the ground state of an odd-$A$ nucleus \cite{Schunck10}. Moreover, 
 since  ground-state configurations of odd-$A$  nuclei internally break the time-reversal symmetry, poorly known  time-odd terms of the EDF must be considered. Although some of the time-odd functionals are constrained through the local gauge invariance of the EDF \cite{PhysRevC.52.1827}, the optimization of the unconstrained time-odd coupling constants has barely  started \cite{PhysRevC.93.014304}.
Consequently,  the precision of the nuclear EDF for odd-$A$ systems is not as good as that for even-even systems. It is thus desirable to constrain the pairing EDF based on experimental data involving even-even systems only.

{\it Objectives}.---In this Letter we assess the performance of nuclear DFT  for pairing-rotational bands in even-even nuclei, both semimagic and doubly-open-shell systems. We study  pairing-rotational moments of inertia and assess their validity as indicators of nucleonic pairing. We check the sensitivity  of pairing rotations in neutron-rich nuclei on the density dependence of the pairing functional.

{\it Definitions}.---The pairing-rotational picture is based on a single intrinsic ``deformed'' one-body field in a gauge space.
The ground-state energy of a system with $N/2$ fermionic pairs can be  expanded up to the second order in the particle number with respect to a reference system with particle number $N_0$ \cite{Bayman69,Beck72,bb77,Brink-Broglia,ZPA275_297},
\begin{align}
  E(N) = E(N_0) + \lambda(N_0)\Delta N + \frac{(\Delta N)^2}{2{\cal J}(N_0)},
  \label{eq:pairrotenergy}
\end{align}
where $\Delta N= N - N_0$,  $\lambda(N_0)=d E/d N|_{N=N_0}$ is the chemical potential, and the second-order term is the pairing-rotational 
energy with the moment of inertia ${\cal J}(N_0)^{-1}=d^2E/d N^2|_{N=N_0}$. In the case of a two-fermion system, 
Eq.~(\ref{eq:pairrotenergy}) can be generalized by considering two coupled pairing-rotational modes. In particular,
when both neutrons and protons exhibit the pair condensate, there exist two NG eigenmodes being linear combinations of the neutron and proton pairing rotations \cite{Marshalek1977,PhysRevC.92.034321}.
(A similar situation in the dense superfluid matter in neutron stars  has recently been discussed in Ref.~\cite{Bedaque2014340}.)
The corresponding rotational energy 
can be written as \cite{ZPA275_297}
\begin{align}\label{Erot}
E_{\rm rot}^{\rm pair}=\sum_{\tau,\tau'=n,p}   \frac{\Delta N_\tau \Delta N_{\tau'}}{2{\cal J}_{\tau \tau'}}, 
\end{align}
where $N_n=N$, $N_p=Z$, $\Delta N_n=N-N_0$, $\Delta N_p=Z-Z_0$, and the tensor
\begin{align}\label{MOI}
\left. {\cal J}_{\tau \tau'}=\frac{\partial N_\tau}{\partial \lambda_{\tau'}}\right|_{\Delta N_{\tau'}=0}=\left.
\left[\frac{\partial ^2E}{\partial N_\tau \partial   N_{\tau'}}\right]^{-1}\right|
_{\Delta N_\tau=\Delta N_{\tau'}=0}
\end{align}
is the pairing-rotational moment of inertia. The tensor ${\cal J}_{\tau \tau'}$
is very sensitive to  pairing correlations. Since it is related to the second derivative of the total energy with respect to particle number, the corresponding Thouless-Valatin (TV) inertia for the NG mode can be 
readily derived by means of
the self-consistent quasiparticle random-phase approximation (QRPA) \cite{Thouless1962211,Ring-Schuck}.

In the region of particle numbers where static pairing dominates, ${\cal J}_{\tau \tau'}$ can be extracted from  experimental two-nucleon separation energies $S_{2n}$ and $S_{2p}$. 
For instance, by taking  $\lambda_{n}(N,Z) =-\frac{1}{4}[S_{2n}(N+2,Z) + S_{2n}(N,Z)]$ the  moments of inertia can be written as
\begin{align}
  {\cal J}^{-1}_{nn}(N,Z) =& \frac{1}{4}\left[S_{2n}(N,Z)- S_{2n}(N+2,Z) \right], \label{MOInn} \\  
  {\cal J}_{np}^{-1}(N,Z) =& \frac{1}{4}\left[S_{2n}(N+2,Z)- S_{2n}(N+2,Z+2) \right]. \label{MOInp}
\end{align}
(The analogous  expressions for  $\lambda_{p}$ and $  {\cal J}_{pp}$
are given in terms of $S_{2p}$.)

{\it Method}.---To compute the TV moments of inertia for pairing rotations we employ
the linear response formalism of nuclear DFT in the finite amplitude method (FAM) \cite{nakatsukasa:024318} variant.
The FAM allows one to handle all the two-quasiparticle states on the QRPA level with a smaller computational cost than that of the traditional matrix formulation of the QRPA.
The TV moment of inertia is given by a response function of the particle number operator at zero frequency. In this study, we follow the FAM formulation of Ref.~\cite{PhysRevC.92.034321} for NG modes.

The computations were  performed with the FAM code  \cite{PhysRevC.84.041305,PhysRevC.91.044323} using the DFT solver \textsc{hfbtho} \cite{Stoitsov20131592}
in a  single-particle basis consisting of
20 harmonic oscillator shells.
We employed the recently developed EDF UNEDF1-HFB  \cite{0954-3899-42-3-034024} that was optimized  at the full Hartree-Fock-Bogoliubov (HFB) level. 
For the pairing energy density  we use the density-dependent ansatz \cite{Dobaczewski2002}
 \begin{align}
   \tilde{\chi}_{\tau}(\bm{r}) = \frac{1}{2} V_0^{\tau} \left[ 1 - \eta \frac{\rho_0(\bm{r})}{\rho_c} \right] |\tilde{\rho}_{\tau}(\bm{r})|^2, \label{eq:pairing}
 \end{align}
where $\tilde{\rho}_\tau$ is the pairing density, $\rho_0$ is the isoscalar density,
 $\rho_c=0.16$ fm$^{-3}$, $V_0^\tau$ is the strength, and $\eta$ is the parameter that controls the density dependence of the pairing interaction.

In  UNEDF1-HFB, mixed-type pairing ($\eta=0.5$) is employed.
To analyze the sensitivity of results on the density dependence of the pairing functional, we also studied volume-type ($\eta=0$) and surface-type ($\eta=1$) pairing  with the  strengths adjusted to reproduce the average neutron pairing gap in $^{120}$Sn and average proton pairing gap in $^{92}$Mo 
assuming the default pairing energy window of 60 MeV.
These nuclei were chosen because the average pairing gaps computed with UNEDF1-HFB are close to the experimental values. The resulting pairing strengths are $V_0^{n}=-146.07$ MeV\,fm$^{3}$ and $V_0^{p}=-161.72$ MeV\,fm$^{3}$ for the volume pairing, and $V_0^{n}=-474.32$ MeV\,fm$^{3}$ and $V_0^{p}=-551.37$ MeV fm$^{3}$ for the surface pairing. 

%%%%%
\begin{figure}[!htb]
 \includegraphics[width=0.8\linewidth]{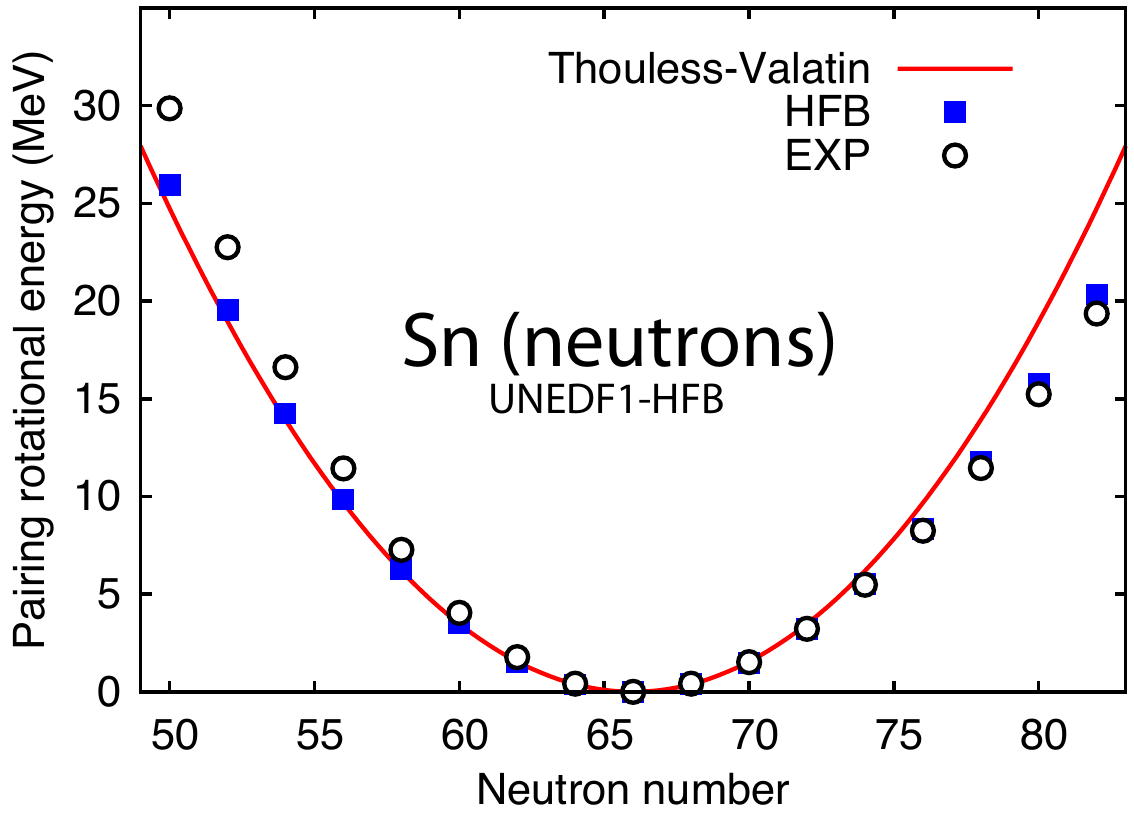}
  \caption{
    Neutron pairing-rotational energy measured from a reference state in $^{116}$Sn. The parabolic expression Eq.~(\ref{eq:pairrotenergy})
with $\lambda_{n}$ and ${\cal J}_{nn}$ evaluated for the reference nucleus is  shown  as a solid  line. 
The HFB (squares) and experimental (circles) values  have been extracted from binding energies according to  $E_{\rm rot}^{\rm pair}=E(N) -E(66) - \lambda_{n}(66)(N-66)$.
    \label{fig:Sn}
    }
\end{figure}
%%%%%%%
{\it Results}.---We start with the classic case of neutron pairing rotations in a semimagic chain of Sn isotopes \cite{Brink-Broglia}.
The theoretical values of the chemical potential and the TV inertia have been computed  for the reference nucleus $^{116}$Sn ($N_0=66$).
As seen in Fig.~\ref{fig:Sn}, the harmonic approximation Eq.~(\ref{eq:pairrotenergy})
works very well in this case; indeed,
the TV pairing inertia agrees  with experiment
even when $N$ is far from $N_0$. This shows that a single intrinsic pairing  field of  $^{116}$Sn explains the binding energy behavior in terms of the dynamics of the NG mode.

\begin{figure*}[!bth]
 \includegraphics[width=1.0\linewidth]{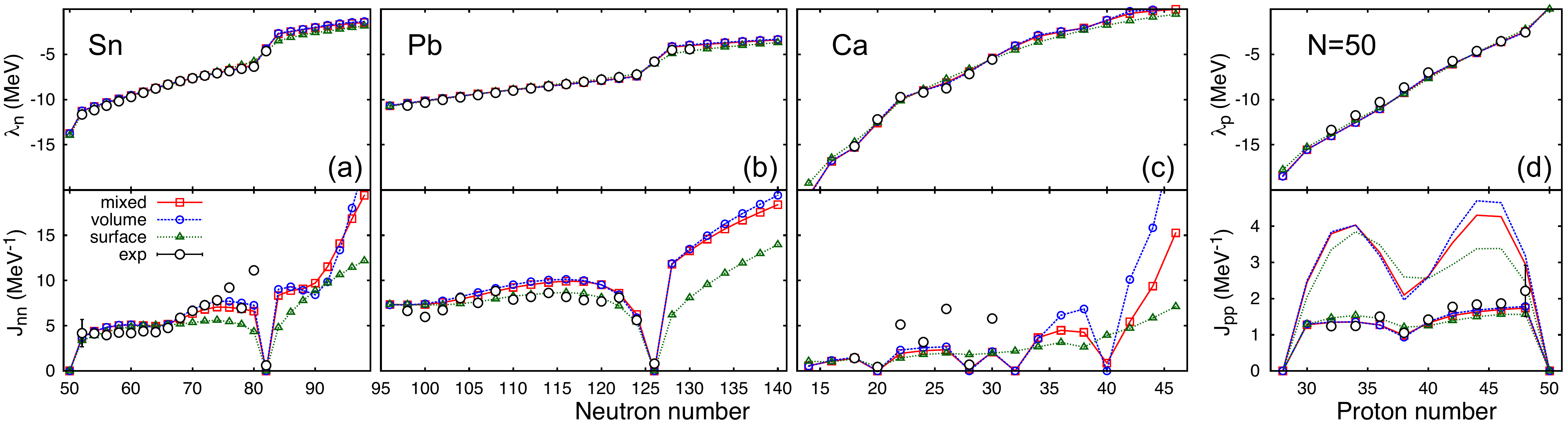}
    \caption{
Chemical potential (top panels) and pairing-rotational moment of inertia (bottom panels) for (a) Sn, (b) Pb, and (c) Ca isotopes, and (d) $N=50$ isotones.
Spherical UNEDF1-HFB solutions with mixed (squares), volume (circles), and surface (triangles) pairing are 
compared to  experimental data from Ref.~\cite{ame2012}.
For $N=50$ isotones, Belyaev moments of inertia are also shown by lines without symbols.
\label{fig:singleclosed}
  }
\end{figure*}

% single-closed nuclei
In general, the higher-order corrections in $\Delta N$ are not  negligible;
in analogy with the angular momentum alignment within a rotational band, a change of the intrinsic structure with  neutron number is expected along a pairing-rotational band. This is  seen in Fig.~\ref{fig:Sn} through the deviation of the HFB values (or experiment)  from parabolic behavior. To account for the changes of the intrinsic pairing field, we carry out systematic FAM+HFB calculations for chains of semimagic nuclei. Figure~\ref{fig:singleclosed} displays  associated chemical potentials and  pairing-rotational moments of inertia. 

The pairing-rotational moments of inertia for Sn and Pb isotopes behave fairly smoothly, and the pairing-rotational picture holds in the medium-mass Ca isotopes.
In general, we see a remarkably good agreement between TV moments of inertia with experiment. The exceptions are weakly paired systems around the magic numbers for which a  transition to the pairing vibrational picture takes place. In such cases, e.g., for $^{130}$Sn and $^{42,46,50}$Ca, the experimental indicator Eq.~(\ref{MOInn}) involves nuclei
for which our HFB calculations predict vanishing pairing. The finite-difference approximation of the second-order derivative is questionable there.

For the doubly-closed-shell nuclei, the theoretical pairing-rotational inertia is zero  as the NG mode is absent due to the vanishing static pairing. 
Moreover, the expression [Eq.~(\ref{MOInn})] for the experimental inertia
${\cal J}_{\tau\tau}$ is proportional to  the inverse of the so-called two-nucleon shell gap indicator $\delta_{2\tau}$ \cite{Bender02,RevModPhys.75.1021}.
This latter quantity has been attributed to the size of the magic gap. 
As it was already noted in Ref.~\cite{Bender02}, the validity of $\delta_{2\tau}$  as a signature of a shell closure is lost in regions where the structure of nuclear ground states is rapidly changing. Based on our results for semimagic
nuclei shown in Fig.~\ref{fig:singleclosed}, we can make an even stronger statement: outside shell closures, the two-nucleon shell gap indicator $\delta_{2\tau}$ has nothing or little to do with the distribution of single-particle energies; it is primarily governed by pairing correlations 
and serves as a good indicator of the gauge symmetry breaking.

%proton pairing
We now study the proton pairing  by investigating pairing rotation in  the $N=50$ isotones. As shown  in Fig.~\ref{fig:singleclosed}(d), 
the proton pairing moments of inertia are smaller than the neutron ones in the similar mass region,
and the agreement with  experiment is excellent. 
In the figure, we also plot the Belyaev moment of inertia \cite{Beliaev1961322}, which does not include the effect of residual correlations at the QRPA level.
As discussed in Ref.~\cite{PhysRevC.92.034321}, the enhancement of the difference between TV and Belyaev proton inertia can be attributed to the Coulomb-induced QRPA correlations. Here we recall that the proton pairing strength required to provide  good agreement with experimental odd-even mass differences is significantly larger than the neutron strength, $V_0^{p}/V_0^{n}\approx 1.1$, and this is consistent with the results of the global survey~\cite{PhysRevC.79.034306}. The large effect of  Coulomb correlations on ${\cal J}_{pp}$, manifesting  itself through the difference between Belyaev and TV proton pairing-rotational inertia, confirms the conclusion of Ref.~\cite{Ang01}
that the Coulomb substantially suppresses proton pairing.

%density dependence
 
The density and momentum  dependence of the pairing functional are not well known
because standard observables probing the pairing channel, such as odd-even mass staggering or moments of inertia of deformed nuclei, show weak sensitivity to details. In this context, the pairing-rotational inertia of single-shell-closed nuclei can serve as a good indicator of the pairing interaction.
The results of calculations for semimagic nuclei in Fig.~\ref{fig:singleclosed}, based on
pairing fitted to experimental odd-even mass differences, are fairly similar for volume-, mixed-, and surface-pairing variants, except for very neutron-rich nuclei where the surface pairing gives appreciably lower values of ${\cal J}_{nn}$. Of particular interest is the behavior of the pairing-rotational inertia in the  very neutron-rich Ca isotopes beyond $^{56}$Ca, where the  pairing functional of volume type yields a $1.5$--$2$ times larger value of ${\cal J}_{nn}$ than the mixed-pairing interaction.  Mass measurements of even-even Ca isotopes beyond $N=36$ will be useful to better constrain the density dependence of the pairing EDF.
Calculations employing the traditional EDFs, such as SLy4 \cite{(Cha98)} and SkM$^*$ \cite{(Bar82)}, show worse agreement with experiment as compared to  UNEDF1-HFB. The latter has been carefully optimized to remove the large systematic errors affecting global binding energy trends \cite{PhysRevC.82.024313}. It is clear, therefore, that to reveal the nature of the pairing functionals through  pairing-rotational inertia one needs to start from the 
well-fitted EDFs in the particle-hole sector.

\begin{figure}[!htb]
 \includegraphics[width=0.9\linewidth]{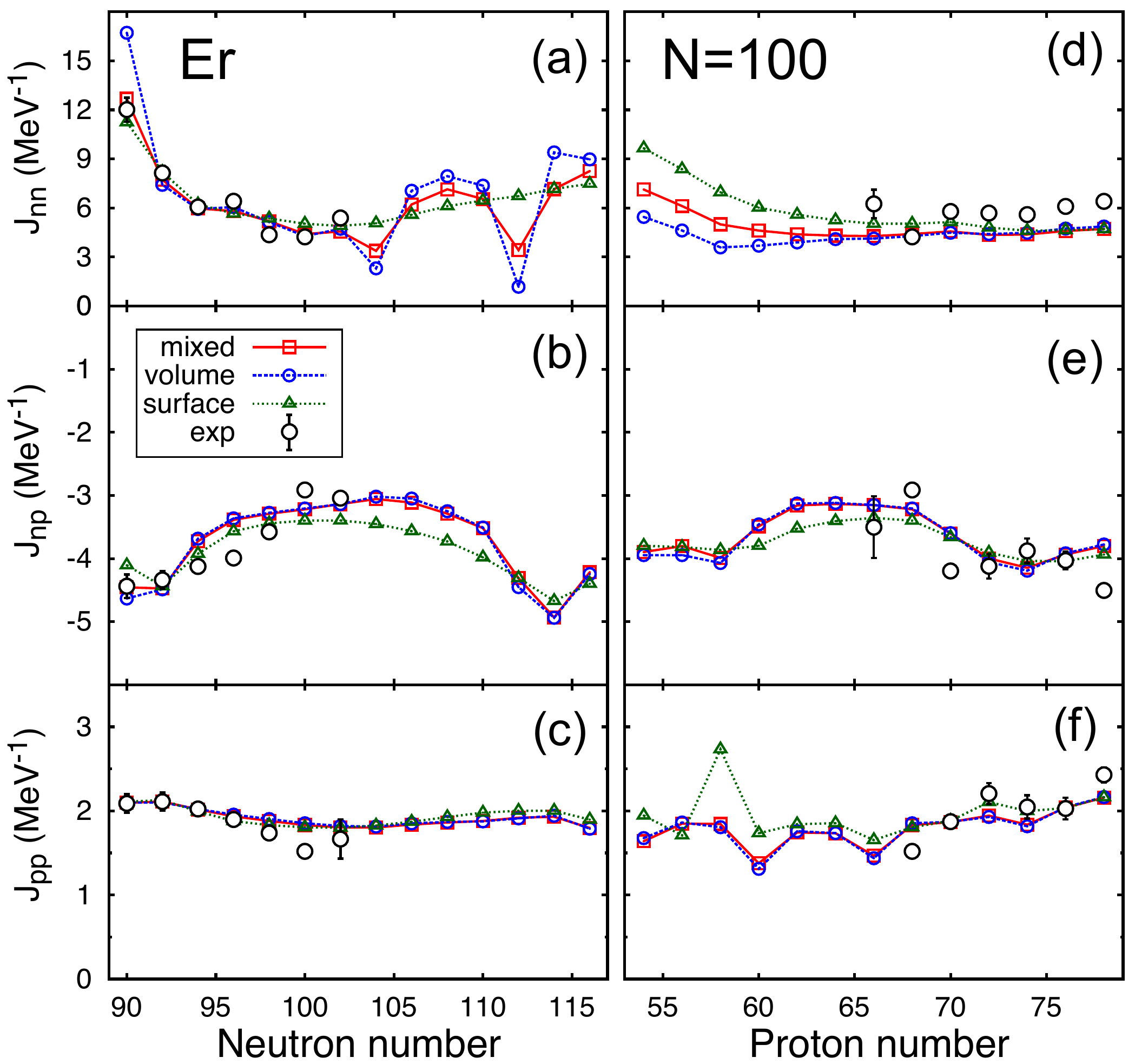}
  \caption{
    Pairing-rotational moments of inertia ${\cal J}_{nn}$ (a,d),
    ${\cal J}_{np}$ (b,e), and ${\cal J}_{pp}$ (c,f) for open-shell
     Er isotopes (left) and $N=100$ isotones (right) exhibiting static  neutron and proton pairing.
    \label{fig:doublyopen}
    }
\end{figure}

% open-open
Finally, we discuss doubly-open-shell nuclei.
When both neutrons and protons show a pair condensate, there exist two NG modes whose  eigenmodes are linear combinations of the neutron and proton pairing rotations \cite{PhysRevC.92.034321}.
In Fig.~\ref{fig:doublyopen}, we show the full  pairing-rotational moment of inertia tensor
for the  Er isotopes and $N=100$ isotones. Both examples are representative of
well-deformed, open-shell nuclei with static neutron and proton pairing.
%%%%
%%%%
Our calculations give excellent  agreement  with  experiment for the full pairing-rotational inertia tensor. In general, the sensitivity to the density dependence of pairing interaction is fairly weak except for very  neutron-rich (proton-deficient) nuclei. 

The off-diagonal moment of inertia ${\cal J}_{np}$  shows quantitative agreement with the experimental data. We emphasize that the example shown in Fig.~\ref{fig:doublyopen} represents  the first systematic calculation of the off-diagonal inertia for two-dimensional pairing rotation, which was seen as a tilted energy kernel in the gauge space in Ref.~\cite{PhysRevC.90.014312}. 
The agreement with  experiment  confirms that the two pairing-rotational NG modes are indeed mixed  through the residual interaction in QRPA. Another interesting aspect of ${\cal J}_{np}$ is that the inverse of this quantity is formally equivalent---up to a trivial shift ($Z\rightarrow Z+2, N\rightarrow N+2$)---to the mass indicator $-\delta V_{pn}$  \cite{Zhang19891}, often referred to as, and interpreted in terms of,  the empirical proton-neutron interaction energy. 
Indeed, in the extreme shell model picture, $\delta V_{pn}$  
represents the net interaction of the last two valence neutrons with the last two valence protons
\cite{Cakirli06,Cakirli06a,Qi2012436,PhysRevC.88.054309}. While the large-scale  superfluid DFT calculations  of $\delta V_{pn}$ generally match the experimental data on the double binding-energy difference [Eq.~(\ref{MOInp})] \cite{PhysRevLett.98.132502}, the direct interpretation of this quantity in terms of the valence proton-neutron interaction is under debate~\cite{PhysRevC.83.064319}. As pointed out in Ref. \cite{PhysRevLett.98.132502}, while the value of $\delta V_{pn}$  averaged over many states (shells) probes the bulk symmetry energy term of the EDF,  the {\it local} behavior of $\delta V_{pn}$  carries  important information about shell effects and many-body correlations.
The relation Eq.~(\ref{MOInp})  between ${\cal J}_{np}$ and $\delta V_{pn}$ sheds new light on  the interpretation of this quantity in doubly-open-shell nuclei; in those nuclei, $\delta V_{pn}$  represents
the simultaneous spontaneous breaking of the neutron {\it and} proton gauge symmetries of the $T=1$ pairing. In this respect, we would question the findings of Ref.~\cite{Qi2012436} that  the pairing energy plays a relatively minor role in understanding of $\delta V_{pn}$.

{\it Conclusions}.---We show that the $T=1$ pairing-rotational moments of inertia of semimagic and doubly-open-shell nuclei  can be described qualitatively within the NG formalism of the broken gauge symmetry. Since the  experimental mass difference relation representing the pairing inertia tensor  is solely based on  binding energies of even-even nuclei, it is an excellent indicator of  nuclear pairing properties. In many respects,  ${\cal J}_{\tau\tau'}$ is superior to other quantities commonly used to inform us about the magnitude of pairing correlations, such as  odd-even mass differences, which involve properties of odd-mass systems that depend on poorly known time-odd fields impacting individual orbits blocked by an odd nucleon.
Furthermore, we demonstrate that the pairing-rotational inertia tensor can be directly expressed in terms of the binding-energy differences $\delta_{2n}$, $\delta_{2p}$,  and $\delta V_{pn}$---all traditionally  regarded as signatures of the valence nucleon properties in the shell model picture. We now propose a unified  interpretation of these quantities in terms of the gauge symmetry breaking associated with the collective $T=1$ pairing phases. Of course, for nuclei close to shell or subshell closures, with weak pairing correlations, the traditional single-particle interpretation is expected to be  more appropriate.

We present the first systematic analysis of the off-diagonal pairing-rotational moments of inertia ${\cal J}_{np}$, and demonstrate the mixing of the neutron and proton pairing-rotational modes in the ground states of open-shell even-even nuclei.  Our analysis of isotopic and isotonic chains  indicates that  the pairing-rotational moments of inertia of neutron-rich nuclei  can be used to constrain the pairing functional of nuclear DFT. 
In this context,
mass measurements of very neutron-rich isotopes are extremely desirable. Theoretically,
clarifying the role of the missing neutron-proton contribution of the $T=1$ pairing functional to ${\cal J}_{np}$
within the  isospin invariant EDF \cite{PhysRevC.69.014316,Sato13,PhysRevC.89.054317} and clarifying the role of various microscopic aspects (effective masses, density dependence, the role of polarization effects, etc.) \cite{Brink-Broglia, Broglia20001,(BroZel)} will be an exciting  subject for future investigations.

\begin{acknowledgments}
This work is supported by the U.S. Department of Energy, Office of Sciences, Office of Nuclear Physics under Awards No.~DE-SC0013365 (Michigan State University) and No. DE-SC0008511 (NUCLEI SciDAC-3).
Numerical calculation was performed at the COMA (PACS-IX) System
at the Center for Computational Sciences, University of Tsukuba.
\end{acknowledgments}
  
\bibliographystyle{apsrev4-1}

\begin{thebibliography}{61}%
\makeatletter
\providecommand \@ifxundefined [1]{%
 \@ifx{#1\undefined}
}%
\providecommand \@ifnum [1]{%
 \ifnum #1\expandafter \@firstoftwo
 \else \expandafter \@secondoftwo
 \fi
}%
\providecommand \@ifx [1]{%
 \ifx #1\expandafter \@firstoftwo
 \else \expandafter \@secondoftwo
 \fi
}%
\providecommand \natexlab [1]{#1}%
\providecommand \enquote  [1]{``#1''}%
\providecommand \bibnamefont  [1]{#1}%
\providecommand \bibfnamefont [1]{#1}%
\providecommand \citenamefont [1]{#1}%
\providecommand \href@noop [0]{\@secondoftwo}%
\providecommand \href [0]{\begingroup \@sanitize@url \@href}%
\providecommand \@href[1]{\@@startlink{#1}\@@href}%
\providecommand \@@href[1]{\endgroup#1\@@endlink}%
\providecommand \@sanitize@url [0]{\catcode `\\12\catcode `\$12\catcode
  `\&12\catcode `\#12\catcode `\^12\catcode `\_12\catcode `\%12\relax}%
\providecommand \@@startlink[1]{}%
\providecommand \@@endlink[0]{}%
\providecommand \url  [0]{\begingroup\@sanitize@url \@url }%
\providecommand \@url [1]{\endgroup\@href {#1}{\urlprefix }}%
\providecommand \urlprefix  [0]{URL }%
\providecommand \Eprint [0]{\href }%
\providecommand \doibase [0]{http://dx.doi.org/}%
\providecommand \selectlanguage [0]{\@gobble}%
\providecommand \bibinfo  [0]{\@secondoftwo}%
\providecommand \bibfield  [0]{\@secondoftwo}%
\providecommand \translation [1]{[#1]}%
\providecommand \BibitemOpen [0]{}%
\providecommand \bibitemStop [0]{}%
\providecommand \bibitemNoStop [0]{.\EOS\space}%
\providecommand \EOS [0]{\spacefactor3000\relax}%
\providecommand \BibitemShut  [1]{\csname bibitem#1\endcsname}%
\let\auto@bib@innerbib\@empty
%</preamble>
\bibitem [{\citenamefont {Anderson}(1958)}]{PhysRev.112.1900}%
  \BibitemOpen
  \bibfield  {author} {\bibinfo {author} {\bibfnamefont {P.~W.}\ \bibnamefont
  {Anderson}},\ }\href {\doibase 10.1103/PhysRev.112.1900} {\bibfield
  {journal} {\bibinfo  {journal} {Phys. Rev.}\ }\textbf {\bibinfo {volume}
  {112}},\ \bibinfo {pages} {1900} (\bibinfo {year} {1958})}\BibitemShut
  {NoStop}%
\bibitem [{\citenamefont {Nambu}(1960)}]{PhysRev.117.648}%
  \BibitemOpen
  \bibfield  {author} {\bibinfo {author} {\bibfnamefont {Y.}~\bibnamefont
  {Nambu}},\ }\href {\doibase 10.1103/PhysRev.117.648} {\bibfield  {journal}
  {\bibinfo  {journal} {Phys. Rev.}\ }\textbf {\bibinfo {volume} {117}},\
  \bibinfo {pages} {648} (\bibinfo {year} {1960})}\BibitemShut {NoStop}%
\bibitem [{\citenamefont {Goldstone}(1961)}]{INC_19_154}%
  \BibitemOpen
  \bibfield  {author} {\bibinfo {author} {\bibfnamefont {J.}~\bibnamefont
  {Goldstone}},\ }\href {\doibase 10.1007/BF02812722} {\bibfield  {journal}
  {\bibinfo  {journal} {Il Nuovo Cimento}\ }\textbf {\bibinfo {volume} {19}},\
  \bibinfo {pages} {154} (\bibinfo {year} {1961})}\BibitemShut {NoStop}%
\bibitem [{\citenamefont {Bes}\ and\ \citenamefont
  {Kurchan}(1990)}]{bes-kurchan}%
  \BibitemOpen
  \bibfield  {author} {\bibinfo {author} {\bibfnamefont {D.~R.}\ \bibnamefont
  {Bes}}\ and\ \bibinfo {author} {\bibfnamefont {J.}~\bibnamefont {Kurchan}},\
  }\href@noop {} {\emph {\bibinfo {title} {The Treatment of Collective
  Coordinates in Many-Body Systems: An Application of the BRST Invariance}}}\
  (\bibinfo  {publisher} {World Scientific, Singapore},\ \bibinfo {year}
  {1990})\BibitemShut {NoStop}%
\bibitem [{\citenamefont {Bohr}\ and\ \citenamefont
  {Mottelson}(1975)}]{Bohr-MottelsonV2}%
  \BibitemOpen
  \bibfield  {author} {\bibinfo {author} {\bibfnamefont {A.}~\bibnamefont
  {Bohr}}\ and\ \bibinfo {author} {\bibfnamefont {B.~R.}\ \bibnamefont
  {Mottelson}},\ }\href@noop {} {\emph {\bibinfo {title} {Nuclear
  Structure}}},\ Vol.~\bibinfo {volume} {II}\ (\bibinfo  {publisher} {W A.
  Benjamin, Reading, MA},\ \bibinfo {year} {1975})\BibitemShut {NoStop}%
\bibitem [{\citenamefont {Reinhard}\ and\ \citenamefont
  {Otten}(1984)}]{PGOtten}%
  \BibitemOpen
  \bibfield  {author} {\bibinfo {author} {\bibfnamefont {P.-G.}\ \bibnamefont
  {Reinhard}}\ and\ \bibinfo {author} {\bibfnamefont {E.~W.}\ \bibnamefont
  {Otten}},\ }\href {\doibase 10.1016/0375-9474(84)90437-8} {\bibfield
  {journal} {\bibinfo  {journal} {Nucl. Phys. A}\ }\textbf {\bibinfo {volume}
  {420}},\ \bibinfo {pages} {173} (\bibinfo {year} {1984})}\BibitemShut
  {NoStop}%
\bibitem [{\citenamefont {Nazarewicz}(1994)}]{Nazarewicz94}%
  \BibitemOpen
  \bibfield  {author} {\bibinfo {author} {\bibfnamefont {W.}~\bibnamefont
  {Nazarewicz}},\ }\href {\doibase 10.1016/0375-9474(94)90037-X} {\bibfield
  {journal} {\bibinfo  {journal} {Nucl. Phys. A}\ }\textbf {\bibinfo {volume}
  {574}},\ \bibinfo {pages} {27 } (\bibinfo {year} {1994})}\BibitemShut
  {NoStop}%
\bibitem [{\citenamefont {Frauendorf}(2001)}]{Frauendorf01}%
  \BibitemOpen
  \bibfield  {author} {\bibinfo {author} {\bibfnamefont {S.}~\bibnamefont
  {Frauendorf}},\ }\href {\doibase 10.1103/RevModPhys.73.463} {\bibfield
  {journal} {\bibinfo  {journal} {Rev. Mod. Phys.}\ }\textbf {\bibinfo {volume}
  {73}},\ \bibinfo {pages} {463} (\bibinfo {year} {2001})}\BibitemShut
  {NoStop}%
\bibitem [{\citenamefont {Bes}\ and\ \citenamefont {Broglia}(1966)}]{bb66}%
  \BibitemOpen
  \bibfield  {author} {\bibinfo {author} {\bibfnamefont {D.~R.}\ \bibnamefont
  {Bes}}\ and\ \bibinfo {author} {\bibfnamefont {R.~A.}\ \bibnamefont
  {Broglia}},\ }\href {\doibase 10.1016/0029-5582(66)90090-3} {\bibfield
  {journal} {\bibinfo  {journal} {Nucl. Phys.}\ }\textbf {\bibinfo {volume}
  {80}},\ \bibinfo {pages} {289 } (\bibinfo {year} {1966})}\BibitemShut
  {NoStop}%
\bibitem [{\citenamefont {Bayman}\ \emph {et~al.}(1969)\citenamefont {Bayman},
  \citenamefont {Bes},\ and\ \citenamefont {Broglia}}]{Bayman69}%
  \BibitemOpen
  \bibfield  {author} {\bibinfo {author} {\bibfnamefont {B.~F.}\ \bibnamefont
  {Bayman}}, \bibinfo {author} {\bibfnamefont {D.~R.}\ \bibnamefont {Bes}}, \
  and\ \bibinfo {author} {\bibfnamefont {R.~A.}\ \bibnamefont {Broglia}},\
  }\href {\doibase 10.1103/PhysRevLett.23.1299} {\bibfield  {journal} {\bibinfo
   {journal} {Phys. Rev. Lett.}\ }\textbf {\bibinfo {volume} {23}},\ \bibinfo
  {pages} {1299} (\bibinfo {year} {1969})}\BibitemShut {NoStop}%
\bibitem [{\citenamefont {Bes}\ and\ \citenamefont {Broglia}(1977)}]{bb77}%
  \BibitemOpen
  \bibfield  {author} {\bibinfo {author} {\bibfnamefont {D.~R.}\ \bibnamefont
  {Bes}}\ and\ \bibinfo {author} {\bibfnamefont {R.~A.}\ \bibnamefont
  {Broglia}},\ }in\ \href {https://books.google.com/books?id=6SkFAQAAIAAJ}
  {\emph {\bibinfo {booktitle} {Proceedings of the International School of
  Physics ``E. Fermi'', Course LXIX}}},\ \bibinfo {editor} {edited by\ \bibinfo
  {editor} {\bibfnamefont {A.}~\bibnamefont {Bohr}}\ and\ \bibinfo {editor}
  {\bibfnamefont {R.~A.}\ \bibnamefont {Broglia}}}\ (\bibinfo  {publisher}
  {North Holland, Amsterdam},\ \bibinfo {year} {1977})\ p.~\bibinfo {pages}
  {55}\BibitemShut {NoStop}%
\bibitem [{\citenamefont {Brink}\ and\ \citenamefont
  {Broglia}(2005)}]{Brink-Broglia}%
  \BibitemOpen
  \bibfield  {author} {\bibinfo {author} {\bibfnamefont {D.~M.}\ \bibnamefont
  {Brink}}\ and\ \bibinfo {author} {\bibfnamefont {R.~A.}\ \bibnamefont
  {Broglia}},\ }\href@noop {} {\emph {\bibinfo {title} {Nuclear Superfluidity,
  Pairing in Finite Systems}}}\ (\bibinfo  {publisher} {Cambridge University
  Press, Cambridge, England},\ \bibinfo {year} {2005})\BibitemShut {NoStop}%
\bibitem [{\citenamefont {Broglia}\ \emph {et~al.}(2000)\citenamefont
  {Broglia}, \citenamefont {Terasaki},\ and\ \citenamefont
  {Giovanardi}}]{Broglia20001}%
  \BibitemOpen
  \bibfield  {author} {\bibinfo {author} {\bibfnamefont {R.~A.}\ \bibnamefont
  {Broglia}}, \bibinfo {author} {\bibfnamefont {J.}~\bibnamefont {Terasaki}}, \
  and\ \bibinfo {author} {\bibfnamefont {N.}~\bibnamefont {Giovanardi}},\
  }\href {\doibase 10.1016/S0370-1573(00)00046-6} {\bibfield  {journal}
  {\bibinfo  {journal} {Phys. Rep.}\ }\textbf {\bibinfo {volume} {335}},\
  \bibinfo {pages} {1 } (\bibinfo {year} {2000})}\BibitemShut {NoStop}%
\bibitem [{\citenamefont {von Oertzen}\ and\ \citenamefont
  {Vitturi}(2001)}]{Oertzen01}%
  \BibitemOpen
  \bibfield  {author} {\bibinfo {author} {\bibfnamefont {W.}~\bibnamefont {von
  Oertzen}}\ and\ \bibinfo {author} {\bibfnamefont {A.}~\bibnamefont
  {Vitturi}},\ }\href {http://stacks.iop.org/0034-4885/64/i=10/a=202}
  {\bibfield  {journal} {\bibinfo  {journal} {Rep. Prog. Phys.}\ }\textbf
  {\bibinfo {volume} {64}},\ \bibinfo {pages} {1247} (\bibinfo {year}
  {2001})}\BibitemShut {NoStop}%
\bibitem [{\citenamefont {Shimoyama}\ and\ \citenamefont
  {Matsuo}(2011)}]{PhysRevC.84.044317}%
  \BibitemOpen
  \bibfield  {author} {\bibinfo {author} {\bibfnamefont {H.}~\bibnamefont
  {Shimoyama}}\ and\ \bibinfo {author} {\bibfnamefont {M.}~\bibnamefont
  {Matsuo}},\ }\href {\doibase 10.1103/PhysRevC.84.044317} {\bibfield
  {journal} {\bibinfo  {journal} {Phys. Rev. C}\ }\textbf {\bibinfo {volume}
  {84}},\ \bibinfo {pages} {044317} (\bibinfo {year} {2011})}\BibitemShut
  {NoStop}%
\bibitem [{\citenamefont {Clark}\ \emph {et~al.}(2006)\citenamefont {Clark},
  \citenamefont {Macchiavelli}, \citenamefont {Fortunato},\ and\ \citenamefont
  {Kr\"ucken}}]{PhysRevLett.96.032501}%
  \BibitemOpen
  \bibfield  {author} {\bibinfo {author} {\bibfnamefont {R.~M.}\ \bibnamefont
  {Clark}}, \bibinfo {author} {\bibfnamefont {A.~O.}\ \bibnamefont
  {Macchiavelli}}, \bibinfo {author} {\bibfnamefont {L.}~\bibnamefont
  {Fortunato}}, \ and\ \bibinfo {author} {\bibfnamefont {R.}~\bibnamefont
  {Kr\"ucken}},\ }\href {\doibase 10.1103/PhysRevLett.96.032501} {\bibfield
  {journal} {\bibinfo  {journal} {Phys. Rev. Lett.}\ }\textbf {\bibinfo
  {volume} {96}},\ \bibinfo {pages} {032501} (\bibinfo {year}
  {2006})}\BibitemShut {NoStop}%
\bibitem [{\citenamefont {Potel}\ \emph {et~al.}(2011)\citenamefont {Potel},
  \citenamefont {Barranco}, \citenamefont {Marini}, \citenamefont {Idini},
  \citenamefont {Vigezzi},\ and\ \citenamefont
  {Broglia}}]{PhysRevLett.107.092501}%
  \BibitemOpen
  \bibfield  {author} {\bibinfo {author} {\bibfnamefont {G.}~\bibnamefont
  {Potel}}, \bibinfo {author} {\bibfnamefont {F.}~\bibnamefont {Barranco}},
  \bibinfo {author} {\bibfnamefont {F.}~\bibnamefont {Marini}}, \bibinfo
  {author} {\bibfnamefont {A.}~\bibnamefont {Idini}}, \bibinfo {author}
  {\bibfnamefont {E.}~\bibnamefont {Vigezzi}}, \ and\ \bibinfo {author}
  {\bibfnamefont {R.~A.}\ \bibnamefont {Broglia}},\ }\href {\doibase
  10.1103/PhysRevLett.107.092501} {\bibfield  {journal} {\bibinfo  {journal}
  {Phys. Rev. Lett.}\ }\textbf {\bibinfo {volume} {107}},\ \bibinfo {pages}
  {092501} (\bibinfo {year} {2011})},\ \bibinfo {note} {{\bf 108}, 069904
  (2012)}\BibitemShut {NoStop}%
\bibitem [{\citenamefont {Potel}\ \emph {et~al.}(2013)\citenamefont {Potel},
  \citenamefont {Idini}, \citenamefont {Barranco}, \citenamefont {Vigezzi},\
  and\ \citenamefont {Broglia}}]{PhysRevC.87.054321}%
  \BibitemOpen
  \bibfield  {author} {\bibinfo {author} {\bibfnamefont {G.}~\bibnamefont
  {Potel}}, \bibinfo {author} {\bibfnamefont {A.}~\bibnamefont {Idini}},
  \bibinfo {author} {\bibfnamefont {F.}~\bibnamefont {Barranco}}, \bibinfo
  {author} {\bibfnamefont {E.}~\bibnamefont {Vigezzi}}, \ and\ \bibinfo
  {author} {\bibfnamefont {R.~A.}\ \bibnamefont {Broglia}},\ }\href {\doibase
  10.1103/PhysRevC.87.054321} {\bibfield  {journal} {\bibinfo  {journal} {Phys.
  Rev. C}\ }\textbf {\bibinfo {volume} {87}},\ \bibinfo {pages} {054321}
  (\bibinfo {year} {2013})}\BibitemShut {NoStop}%
\bibitem [{\citenamefont {Perli\ifmmode~\acute{n}\else \'{n}\fi{}ska}\ \emph
  {et~al.}(2004)\citenamefont {Perli\ifmmode~\acute{n}\else \'{n}\fi{}ska},
  \citenamefont {Rohozi\ifmmode~\acute{n}\else \'{n}\fi{}ski}, \citenamefont
  {Dobaczewski},\ and\ \citenamefont {Nazarewicz}}]{PhysRevC.69.014316}%
  \BibitemOpen
  \bibfield  {author} {\bibinfo {author} {\bibfnamefont {E.}~\bibnamefont
  {Perli\ifmmode~\acute{n}\else \'{n}\fi{}ska}}, \bibinfo {author}
  {\bibfnamefont {S.~G.}\ \bibnamefont {Rohozi\ifmmode~\acute{n}\else
  \'{n}\fi{}ski}}, \bibinfo {author} {\bibfnamefont {J.}~\bibnamefont
  {Dobaczewski}}, \ and\ \bibinfo {author} {\bibfnamefont {W.}~\bibnamefont
  {Nazarewicz}},\ }\href {\doibase 10.1103/PhysRevC.69.014316} {\bibfield
  {journal} {\bibinfo  {journal} {Phys. Rev. C}\ }\textbf {\bibinfo {volume}
  {69}},\ \bibinfo {pages} {014316} (\bibinfo {year} {2004})}\BibitemShut
  {NoStop}%
\bibitem [{\citenamefont {Rohozi{\'n}ski}\ \emph {et~al.}(2010)\citenamefont
  {Rohozi{\'n}ski}, \citenamefont {Dobaczewski},\ and\ \citenamefont
  {Nazarewicz}}]{Roh10}%
  \BibitemOpen
  \bibfield  {author} {\bibinfo {author} {\bibfnamefont {S.~G.}\ \bibnamefont
  {Rohozi{\'n}ski}}, \bibinfo {author} {\bibfnamefont {J.}~\bibnamefont
  {Dobaczewski}}, \ and\ \bibinfo {author} {\bibfnamefont {W.}~\bibnamefont
  {Nazarewicz}},\ }\href {\doibase 10.1103/PhysRevC.81.014313} {\bibfield
  {journal} {\bibinfo  {journal} {Phys. Rev. C}\ }\textbf {\bibinfo {volume}
  {81}},\ \bibinfo {pages} {014313} (\bibinfo {year} {2010})}\BibitemShut
  {NoStop}%
\bibitem [{\citenamefont {Bjerregaard}\ \emph {et~al.}(1968)\citenamefont
  {Bjerregaard}, \citenamefont {Hansen}, \citenamefont {Nathan}, \citenamefont
  {Vistisen}, \citenamefont {Chapman},\ and\ \citenamefont
  {Hinds}}]{Bjerregaard19681}%
  \BibitemOpen
  \bibfield  {author} {\bibinfo {author} {\bibfnamefont {J.}~\bibnamefont
  {Bjerregaard}}, \bibinfo {author} {\bibfnamefont {O.}~\bibnamefont {Hansen}},
  \bibinfo {author} {\bibfnamefont {O.}~\bibnamefont {Nathan}}, \bibinfo
  {author} {\bibfnamefont {L.}~\bibnamefont {Vistisen}}, \bibinfo {author}
  {\bibfnamefont {R.}~\bibnamefont {Chapman}}, \ and\ \bibinfo {author}
  {\bibfnamefont {S.}~\bibnamefont {Hinds}},\ }\href {\doibase
  http://dx.doi.org/10.1016/0375-9474(68)90678-7} {\bibfield  {journal}
  {\bibinfo  {journal} {Nucl. Phys. A}\ }\textbf {\bibinfo {volume} {110}},\
  \bibinfo {pages} {1} (\bibinfo {year} {1968})}\BibitemShut {NoStop}%
\bibitem [{\citenamefont {Kortelainen}\ \emph {et~al.}(2010)\citenamefont
  {Kortelainen}, \citenamefont {Lesinski}, \citenamefont {Mor\'e},
  \citenamefont {Nazarewicz}, \citenamefont {Sarich}, \citenamefont {Schunck},
  \citenamefont {Stoitsov},\ and\ \citenamefont {Wild}}]{PhysRevC.82.024313}%
  \BibitemOpen
  \bibfield  {author} {\bibinfo {author} {\bibfnamefont {M.}~\bibnamefont
  {Kortelainen}}, \bibinfo {author} {\bibfnamefont {T.}~\bibnamefont
  {Lesinski}}, \bibinfo {author} {\bibfnamefont {J.}~\bibnamefont {Mor\'e}},
  \bibinfo {author} {\bibfnamefont {W.}~\bibnamefont {Nazarewicz}}, \bibinfo
  {author} {\bibfnamefont {J.}~\bibnamefont {Sarich}}, \bibinfo {author}
  {\bibfnamefont {N.}~\bibnamefont {Schunck}}, \bibinfo {author} {\bibfnamefont
  {M.~V.}\ \bibnamefont {Stoitsov}}, \ and\ \bibinfo {author} {\bibfnamefont
  {S.}~\bibnamefont {Wild}},\ }\href {\doibase 10.1103/PhysRevC.82.024313}
  {\bibfield  {journal} {\bibinfo  {journal} {Phys. Rev. C}\ }\textbf {\bibinfo
  {volume} {82}},\ \bibinfo {pages} {024313} (\bibinfo {year}
  {2010})}\BibitemShut {NoStop}%
\bibitem [{\citenamefont {Kortelainen}\ \emph {et~al.}(2012)\citenamefont
  {Kortelainen}, \citenamefont {McDonnell}, \citenamefont {Nazarewicz},
  \citenamefont {Reinhard}, \citenamefont {Sarich}, \citenamefont {Schunck},
  \citenamefont {Stoitsov},\ and\ \citenamefont {Wild}}]{PhysRevC.85.024304}%
  \BibitemOpen
  \bibfield  {author} {\bibinfo {author} {\bibfnamefont {M.}~\bibnamefont
  {Kortelainen}}, \bibinfo {author} {\bibfnamefont {J.}~\bibnamefont
  {McDonnell}}, \bibinfo {author} {\bibfnamefont {W.}~\bibnamefont
  {Nazarewicz}}, \bibinfo {author} {\bibfnamefont {P.-G.}\ \bibnamefont
  {Reinhard}}, \bibinfo {author} {\bibfnamefont {J.}~\bibnamefont {Sarich}},
  \bibinfo {author} {\bibfnamefont {N.}~\bibnamefont {Schunck}}, \bibinfo
  {author} {\bibfnamefont {M.~V.}\ \bibnamefont {Stoitsov}}, \ and\ \bibinfo
  {author} {\bibfnamefont {S.~M.}\ \bibnamefont {Wild}},\ }\href {\doibase
  10.1103/PhysRevC.85.024304} {\bibfield  {journal} {\bibinfo  {journal} {Phys.
  Rev. C}\ }\textbf {\bibinfo {volume} {85}},\ \bibinfo {pages} {024304}
  (\bibinfo {year} {2012})}\BibitemShut {NoStop}%
\bibitem [{\citenamefont {Kortelainen}\ \emph {et~al.}(2014)\citenamefont
  {Kortelainen}, \citenamefont {McDonnell}, \citenamefont {Nazarewicz},
  \citenamefont {Olsen}, \citenamefont {Reinhard}, \citenamefont {Sarich},
  \citenamefont {Schunck}, \citenamefont {Wild}, \citenamefont {Davesne},
  \citenamefont {Erler},\ and\ \citenamefont {Pastore}}]{PhysRevC.89.054314}%
  \BibitemOpen
  \bibfield  {author} {\bibinfo {author} {\bibfnamefont {M.}~\bibnamefont
  {Kortelainen}}, \bibinfo {author} {\bibfnamefont {J.}~\bibnamefont
  {McDonnell}}, \bibinfo {author} {\bibfnamefont {W.}~\bibnamefont
  {Nazarewicz}}, \bibinfo {author} {\bibfnamefont {E.}~\bibnamefont {Olsen}},
  \bibinfo {author} {\bibfnamefont {P.-G.}\ \bibnamefont {Reinhard}}, \bibinfo
  {author} {\bibfnamefont {J.}~\bibnamefont {Sarich}}, \bibinfo {author}
  {\bibfnamefont {N.}~\bibnamefont {Schunck}}, \bibinfo {author} {\bibfnamefont
  {S.~M.}\ \bibnamefont {Wild}}, \bibinfo {author} {\bibfnamefont
  {D.}~\bibnamefont {Davesne}}, \bibinfo {author} {\bibfnamefont
  {J.}~\bibnamefont {Erler}}, \ and\ \bibinfo {author} {\bibfnamefont
  {A.}~\bibnamefont {Pastore}},\ }\href {\doibase 10.1103/PhysRevC.89.054314}
  {\bibfield  {journal} {\bibinfo  {journal} {Phys. Rev. C}\ }\textbf {\bibinfo
  {volume} {89}},\ \bibinfo {pages} {054314} (\bibinfo {year}
  {2014})}\BibitemShut {NoStop}%
\bibitem [{\citenamefont {Satu\l{}a}\ \emph {et~al.}(1998)\citenamefont
  {Satu\l{}a}, \citenamefont {Dobaczewski},\ and\ \citenamefont
  {Nazarewicz}}]{Sat98}%
  \BibitemOpen
  \bibfield  {author} {\bibinfo {author} {\bibfnamefont {W.}~\bibnamefont
  {Satu\l{}a}}, \bibinfo {author} {\bibfnamefont {J.}~\bibnamefont
  {Dobaczewski}}, \ and\ \bibinfo {author} {\bibfnamefont {W.}~\bibnamefont
  {Nazarewicz}},\ }\href {\doibase 10.1103/PhysRevLett.81.3599} {\bibfield
  {journal} {\bibinfo  {journal} {Phys. Rev. Lett.}\ }\textbf {\bibinfo
  {volume} {81}},\ \bibinfo {pages} {3599} (\bibinfo {year}
  {1998})}\BibitemShut {NoStop}%
\bibitem [{\citenamefont {Duguet}\ \emph {et~al.}(2001)\citenamefont {Duguet},
  \citenamefont {Bonche}, \citenamefont {Heenen},\ and\ \citenamefont
  {Meyer}}]{Dug01}%
  \BibitemOpen
  \bibfield  {author} {\bibinfo {author} {\bibfnamefont {T.}~\bibnamefont
  {Duguet}}, \bibinfo {author} {\bibfnamefont {P.}~\bibnamefont {Bonche}},
  \bibinfo {author} {\bibfnamefont {P.-H.}\ \bibnamefont {Heenen}}, \ and\
  \bibinfo {author} {\bibfnamefont {J.}~\bibnamefont {Meyer}},\ }\href
  {\doibase 10.1103/PhysRevC.65.014311} {\bibfield  {journal} {\bibinfo
  {journal} {Phys. Rev. C}\ }\textbf {\bibinfo {volume} {65}},\ \bibinfo
  {pages} {014311} (\bibinfo {year} {2001})}\BibitemShut {NoStop}%
\bibitem [{\citenamefont {Bertsch}\ \emph {et~al.}(2009)\citenamefont
  {Bertsch}, \citenamefont {Bertulani}, \citenamefont {Nazarewicz},
  \citenamefont {Schunck},\ and\ \citenamefont
  {Stoitsov}}]{PhysRevC.79.034306}%
  \BibitemOpen
  \bibfield  {author} {\bibinfo {author} {\bibfnamefont {G.~F.}\ \bibnamefont
  {Bertsch}}, \bibinfo {author} {\bibfnamefont {C.~A.}\ \bibnamefont
  {Bertulani}}, \bibinfo {author} {\bibfnamefont {W.}~\bibnamefont
  {Nazarewicz}}, \bibinfo {author} {\bibfnamefont {N.}~\bibnamefont {Schunck}},
  \ and\ \bibinfo {author} {\bibfnamefont {M.~V.}\ \bibnamefont {Stoitsov}},\
  }\href {\doibase 10.1103/PhysRevC.79.034306} {\bibfield  {journal} {\bibinfo
  {journal} {Phys. Rev. C}\ }\textbf {\bibinfo {volume} {79}},\ \bibinfo
  {pages} {034306} (\bibinfo {year} {2009})}\BibitemShut {NoStop}%
\bibitem [{\citenamefont {Schunck}\ \emph {et~al.}(2010)\citenamefont
  {Schunck}, \citenamefont {Dobaczewski}, \citenamefont {McDonnell},
  \citenamefont {Mor\'e}, \citenamefont {Nazarewicz}, \citenamefont {Sarich},\
  and\ \citenamefont {Stoitsov}}]{Schunck10}%
  \BibitemOpen
  \bibfield  {author} {\bibinfo {author} {\bibfnamefont {N.}~\bibnamefont
  {Schunck}}, \bibinfo {author} {\bibfnamefont {J.}~\bibnamefont
  {Dobaczewski}}, \bibinfo {author} {\bibfnamefont {J.}~\bibnamefont
  {McDonnell}}, \bibinfo {author} {\bibfnamefont {J.}~\bibnamefont {Mor\'e}},
  \bibinfo {author} {\bibfnamefont {W.}~\bibnamefont {Nazarewicz}}, \bibinfo
  {author} {\bibfnamefont {J.}~\bibnamefont {Sarich}}, \ and\ \bibinfo {author}
  {\bibfnamefont {M.~V.}\ \bibnamefont {Stoitsov}},\ }\href {\doibase
  10.1103/PhysRevC.81.024316} {\bibfield  {journal} {\bibinfo  {journal} {Phys.
  Rev. C}\ }\textbf {\bibinfo {volume} {81}},\ \bibinfo {pages} {024316}
  (\bibinfo {year} {2010})}\BibitemShut {NoStop}%
\bibitem [{\citenamefont {Dobaczewski}\ and\ \citenamefont
  {Dudek}(1995)}]{PhysRevC.52.1827}%
  \BibitemOpen
  \bibfield  {author} {\bibinfo {author} {\bibfnamefont {J.}~\bibnamefont
  {Dobaczewski}}\ and\ \bibinfo {author} {\bibfnamefont {J.}~\bibnamefont
  {Dudek}},\ }\href {\doibase 10.1103/PhysRevC.52.1827} {\bibfield  {journal}
  {\bibinfo  {journal} {Phys. Rev. C}\ }\textbf {\bibinfo {volume} {52}},\
  \bibinfo {pages} {1827} (\bibinfo {year} {1995})}\BibitemShut {NoStop}%
\bibitem [{\citenamefont {Mustonen}\ and\ \citenamefont
  {Engel}(2016)}]{PhysRevC.93.014304}%
  \BibitemOpen
  \bibfield  {author} {\bibinfo {author} {\bibfnamefont {M.~T.}\ \bibnamefont
  {Mustonen}}\ and\ \bibinfo {author} {\bibfnamefont {J.}~\bibnamefont
  {Engel}},\ }\href {\doibase 10.1103/PhysRevC.93.014304} {\bibfield  {journal}
  {\bibinfo  {journal} {Phys. Rev. C}\ }\textbf {\bibinfo {volume} {93}},\
  \bibinfo {pages} {014304} (\bibinfo {year} {2016})}\BibitemShut {NoStop}%
\bibitem [{\citenamefont {Beck}\ \emph {et~al.}(1972)\citenamefont {Beck},
  \citenamefont {Kleber},\ and\ \citenamefont {Schmidt}}]{Beck72}%
  \BibitemOpen
  \bibfield  {author} {\bibinfo {author} {\bibfnamefont {R.}~\bibnamefont
  {Beck}}, \bibinfo {author} {\bibfnamefont {M.}~\bibnamefont {Kleber}}, \ and\
  \bibinfo {author} {\bibfnamefont {H.}~\bibnamefont {Schmidt}},\ }\href
  {\doibase 10.1007/BF01386946} {\bibfield  {journal} {\bibinfo  {journal} {Z.
  Phys.}\ }\textbf {\bibinfo {volume} {250}},\ \bibinfo {pages} {155} (\bibinfo
  {year} {1972})}\BibitemShut {NoStop}%
\bibitem [{\citenamefont {Krappe}(1975)}]{ZPA275_297}%
  \BibitemOpen
  \bibfield  {author} {\bibinfo {author} {\bibfnamefont {H.~J.}\ \bibnamefont
  {Krappe}},\ }\href {\doibase 10.1007/BF01409299} {\bibfield  {journal}
  {\bibinfo  {journal} {Z. Phys. A}\ }\textbf {\bibinfo {volume} {275}},\
  \bibinfo {pages} {297 } (\bibinfo {year} {1975})}\BibitemShut {NoStop}%
\bibitem [{\citenamefont {Marshalek}(1977)}]{Marshalek1977}%
  \BibitemOpen
  \bibfield  {author} {\bibinfo {author} {\bibfnamefont {E.}~\bibnamefont
  {Marshalek}},\ }\href {\doibase 10.1016/0375-9474(77)90461-4} {\bibfield
  {journal} {\bibinfo  {journal} {Nucl. Phys. A}\ }\textbf {\bibinfo {volume}
  {275}},\ \bibinfo {pages} {416 } (\bibinfo {year} {1977})}\BibitemShut
  {NoStop}%
\bibitem [{\citenamefont {Hinohara}(2015)}]{PhysRevC.92.034321}%
  \BibitemOpen
  \bibfield  {author} {\bibinfo {author} {\bibfnamefont {N.}~\bibnamefont
  {Hinohara}},\ }\href {\doibase 10.1103/PhysRevC.92.034321} {\bibfield
  {journal} {\bibinfo  {journal} {Phys. Rev. C}\ }\textbf {\bibinfo {volume}
  {92}},\ \bibinfo {pages} {034321} (\bibinfo {year} {2015})}\BibitemShut
  {NoStop}%
\bibitem [{\citenamefont {Bedaque}\ and\ \citenamefont
  {Reddy}(2014)}]{Bedaque2014340}%
  \BibitemOpen
  \bibfield  {author} {\bibinfo {author} {\bibfnamefont {P.~F.}\ \bibnamefont
  {Bedaque}}\ and\ \bibinfo {author} {\bibfnamefont {S.}~\bibnamefont
  {Reddy}},\ }\href {\doibase http://dx.doi.org/10.1016/j.physletb.2014.06.033}
  {\bibfield  {journal} {\bibinfo  {journal} {Phys. Lett. B}\ }\textbf
  {\bibinfo {volume} {735}},\ \bibinfo {pages} {340 } (\bibinfo {year}
  {2014})}\BibitemShut {NoStop}%
\bibitem [{\citenamefont {Thouless}\ and\ \citenamefont
  {Valatin}(1962)}]{Thouless1962211}%
  \BibitemOpen
  \bibfield  {author} {\bibinfo {author} {\bibfnamefont {D.~J.}\ \bibnamefont
  {Thouless}}\ and\ \bibinfo {author} {\bibfnamefont {J.~G.}\ \bibnamefont
  {Valatin}},\ }\href {\doibase 10.1016/0029-5582(62)90741-1} {\bibfield
  {journal} {\bibinfo  {journal} {Nucl. Phys.}\ }\textbf {\bibinfo {volume}
  {31}},\ \bibinfo {pages} {211 } (\bibinfo {year} {1962})}\BibitemShut
  {NoStop}%
\bibitem [{\citenamefont {Ring}\ and\ \citenamefont
  {Schuck}(2000)}]{Ring-Schuck}%
  \BibitemOpen
  \bibfield  {author} {\bibinfo {author} {\bibfnamefont {P.}~\bibnamefont
  {Ring}}\ and\ \bibinfo {author} {\bibfnamefont {P.}~\bibnamefont {Schuck}},\
  }\href@noop {} {\emph {\bibinfo {title} {The Nuclear Many-Body Problem}}}\
  (\bibinfo  {publisher} {Springer-Verlag, Berlin},\ \bibinfo {year}
  {2000})\BibitemShut {NoStop}%
\bibitem [{\citenamefont {Nakatsukasa}\ \emph {et~al.}(2007)\citenamefont
  {Nakatsukasa}, \citenamefont {Inakura},\ and\ \citenamefont
  {Yabana}}]{nakatsukasa:024318}%
  \BibitemOpen
  \bibfield  {author} {\bibinfo {author} {\bibfnamefont {T.}~\bibnamefont
  {Nakatsukasa}}, \bibinfo {author} {\bibfnamefont {T.}~\bibnamefont
  {Inakura}}, \ and\ \bibinfo {author} {\bibfnamefont {K.}~\bibnamefont
  {Yabana}},\ }\href {\doibase 10.1103/PhysRevC.76.024318} {\bibfield
  {journal} {\bibinfo  {journal} {Phys. Rev. C}\ }\textbf {\bibinfo {volume}
  {76}},\ \bibinfo {eid} {024318} (\bibinfo {year} {2007})}\BibitemShut
  {NoStop}%
\bibitem [{\citenamefont {Stoitsov}\ \emph {et~al.}(2011)\citenamefont
  {Stoitsov}, \citenamefont {Kortelainen}, \citenamefont {Nakatsukasa},
  \citenamefont {Losa},\ and\ \citenamefont {Nazarewicz}}]{PhysRevC.84.041305}%
  \BibitemOpen
  \bibfield  {author} {\bibinfo {author} {\bibfnamefont {M.}~\bibnamefont
  {Stoitsov}}, \bibinfo {author} {\bibfnamefont {M.}~\bibnamefont
  {Kortelainen}}, \bibinfo {author} {\bibfnamefont {T.}~\bibnamefont
  {Nakatsukasa}}, \bibinfo {author} {\bibfnamefont {C.}~\bibnamefont {Losa}}, \
  and\ \bibinfo {author} {\bibfnamefont {W.}~\bibnamefont {Nazarewicz}},\
  }\href {\doibase 10.1103/PhysRevC.84.041305} {\bibfield  {journal} {\bibinfo
  {journal} {Phys. Rev. C}\ }\textbf {\bibinfo {volume} {84}},\ \bibinfo
  {pages} {041305} (\bibinfo {year} {2011})}\BibitemShut {NoStop}%
\bibitem [{\citenamefont {Hinohara}\ \emph {et~al.}(2015)\citenamefont
  {Hinohara}, \citenamefont {Kortelainen}, \citenamefont {Nazarewicz},\ and\
  \citenamefont {Olsen}}]{PhysRevC.91.044323}%
  \BibitemOpen
  \bibfield  {author} {\bibinfo {author} {\bibfnamefont {N.}~\bibnamefont
  {Hinohara}}, \bibinfo {author} {\bibfnamefont {M.}~\bibnamefont
  {Kortelainen}}, \bibinfo {author} {\bibfnamefont {W.}~\bibnamefont
  {Nazarewicz}}, \ and\ \bibinfo {author} {\bibfnamefont {E.}~\bibnamefont
  {Olsen}},\ }\href {\doibase 10.1103/PhysRevC.91.044323} {\bibfield  {journal}
  {\bibinfo  {journal} {Phys. Rev. C}\ }\textbf {\bibinfo {volume} {91}},\
  \bibinfo {pages} {044323} (\bibinfo {year} {2015})}\BibitemShut {NoStop}%
\bibitem [{\citenamefont {Stoitsov}\ \emph {et~al.}(2013)\citenamefont
  {Stoitsov}, \citenamefont {Schunck}, \citenamefont {Kortelainen},
  \citenamefont {Michel}, \citenamefont {Nam}, \citenamefont {Olsen},
  \citenamefont {Sarich},\ and\ \citenamefont {Wild}}]{Stoitsov20131592}%
  \BibitemOpen
  \bibfield  {author} {\bibinfo {author} {\bibfnamefont {M.~V.}\ \bibnamefont
  {Stoitsov}}, \bibinfo {author} {\bibfnamefont {N.}~\bibnamefont {Schunck}},
  \bibinfo {author} {\bibfnamefont {M.}~\bibnamefont {Kortelainen}}, \bibinfo
  {author} {\bibfnamefont {N.}~\bibnamefont {Michel}}, \bibinfo {author}
  {\bibfnamefont {H.}~\bibnamefont {Nam}}, \bibinfo {author} {\bibfnamefont
  {E.}~\bibnamefont {Olsen}}, \bibinfo {author} {\bibfnamefont
  {J.}~\bibnamefont {Sarich}}, \ and\ \bibinfo {author} {\bibfnamefont
  {S.}~\bibnamefont {Wild}},\ }\href {\doibase 10.1016/j.cpc.2013.01.013}
  {\bibfield  {journal} {\bibinfo  {journal} {Comput. Phys. Commun.}\ }\textbf
  {\bibinfo {volume} {184}},\ \bibinfo {pages} {1592} (\bibinfo {year}
  {2013})}\BibitemShut {NoStop}%
\bibitem [{\citenamefont {Schunck}\ \emph {et~al.}(2015)\citenamefont
  {Schunck}, \citenamefont {McDonnell}, \citenamefont {Sarich}, \citenamefont
  {Wild},\ and\ \citenamefont {Higdon}}]{0954-3899-42-3-034024}%
  \BibitemOpen
  \bibfield  {author} {\bibinfo {author} {\bibfnamefont {N.}~\bibnamefont
  {Schunck}}, \bibinfo {author} {\bibfnamefont {J.~D.}\ \bibnamefont
  {McDonnell}}, \bibinfo {author} {\bibfnamefont {J.}~\bibnamefont {Sarich}},
  \bibinfo {author} {\bibfnamefont {S.~M.}\ \bibnamefont {Wild}}, \ and\
  \bibinfo {author} {\bibfnamefont {D.}~\bibnamefont {Higdon}},\ }\href
  {http://stacks.iop.org/0954-3899/42/i=3/a=034024} {\bibfield  {journal}
  {\bibinfo  {journal} {J. Phys. G}\ }\textbf {\bibinfo {volume} {42}},\
  \bibinfo {pages} {034024} (\bibinfo {year} {2015})}\BibitemShut {NoStop}%
\bibitem [{\citenamefont {Dobaczewski}\ \emph {et~al.}(2002)\citenamefont
  {Dobaczewski}, \citenamefont {Nazarewicz},\ and\ \citenamefont
  {Stoitsov}}]{Dobaczewski2002}%
  \BibitemOpen
  \bibfield  {author} {\bibinfo {author} {\bibfnamefont {J.}~\bibnamefont
  {Dobaczewski}}, \bibinfo {author} {\bibfnamefont {W.}~\bibnamefont
  {Nazarewicz}}, \ and\ \bibinfo {author} {\bibfnamefont {M.~V.}\ \bibnamefont
  {Stoitsov}},\ }\href {\doibase 10.1140/epja/i2001-10218-8} {\bibfield
  {journal} {\bibinfo  {journal} {Eur. Phys. J. A}\ }\textbf {\bibinfo {volume}
  {15}},\ \bibinfo {pages} {21} (\bibinfo {year} {2002})}\BibitemShut {NoStop}%
\bibitem [{\citenamefont {Audi}\ \emph {et~al.}(2012)\citenamefont {Audi},
  \citenamefont {Wang}, \citenamefont {Wapstra}, \citenamefont {Kondev},
  \citenamefont {MacCormick}, \citenamefont {Xu},\ and\ \citenamefont
  {Pfeiffer}}]{ame2012}%
  \BibitemOpen
  \bibfield  {author} {\bibinfo {author} {\bibfnamefont {G.}~\bibnamefont
  {Audi}}, \bibinfo {author} {\bibfnamefont {M.}~\bibnamefont {Wang}}, \bibinfo
  {author} {\bibfnamefont {A.~H.}\ \bibnamefont {Wapstra}}, \bibinfo {author}
  {\bibfnamefont {F.~G.}\ \bibnamefont {Kondev}}, \bibinfo {author}
  {\bibfnamefont {M.}~\bibnamefont {MacCormick}}, \bibinfo {author}
  {\bibfnamefont {X.}~\bibnamefont {Xu}}, \ and\ \bibinfo {author}
  {\bibfnamefont {B.}~\bibnamefont {Pfeiffer}},\ }\href
  {http://stacks.iop.org/1674-1137/36/i=12/a=002} {\bibfield  {journal}
  {\bibinfo  {journal} {Chin. Phys. C}\ }\textbf {\bibinfo {volume} {36}},\
  \bibinfo {pages} {1287} (\bibinfo {year} {2012})}\BibitemShut {NoStop}%
\bibitem [{\citenamefont {Bender}\ \emph {et~al.}(2002)\citenamefont {Bender},
  \citenamefont {Cornelius}, \citenamefont {Lalazissis}, \citenamefont
  {Maruhn}, \citenamefont {Nazarewicz},\ and\ \citenamefont
  {Reinhard}}]{Bender02}%
  \BibitemOpen
  \bibfield  {author} {\bibinfo {author} {\bibfnamefont {M.}~\bibnamefont
  {Bender}}, \bibinfo {author} {\bibfnamefont {T.}~\bibnamefont {Cornelius}},
  \bibinfo {author} {\bibfnamefont {G.~A.}\ \bibnamefont {Lalazissis}},
  \bibinfo {author} {\bibfnamefont {J.~A.}\ \bibnamefont {Maruhn}}, \bibinfo
  {author} {\bibfnamefont {W.}~\bibnamefont {Nazarewicz}}, \ and\ \bibinfo
  {author} {\bibfnamefont {P.-G.}\ \bibnamefont {Reinhard}},\ }\href {\doibase
  10.1140/epja/iepja1320} {\bibfield  {journal} {\bibinfo  {journal} {Eur.
  Phys. J. A}\ }\textbf {\bibinfo {volume} {14}},\ \bibinfo {pages} {23}
  (\bibinfo {year} {2002})}\BibitemShut {NoStop}%
\bibitem [{\citenamefont {Lunney}\ \emph {et~al.}(2003)\citenamefont {Lunney},
  \citenamefont {Pearson},\ and\ \citenamefont
  {Thibault}}]{RevModPhys.75.1021}%
  \BibitemOpen
  \bibfield  {author} {\bibinfo {author} {\bibfnamefont {D.}~\bibnamefont
  {Lunney}}, \bibinfo {author} {\bibfnamefont {J.~M.}\ \bibnamefont {Pearson}},
  \ and\ \bibinfo {author} {\bibfnamefont {C.}~\bibnamefont {Thibault}},\
  }\href {\doibase 10.1103/RevModPhys.75.1021} {\bibfield  {journal} {\bibinfo
  {journal} {Rev. Mod. Phys.}\ }\textbf {\bibinfo {volume} {75}},\ \bibinfo
  {pages} {1021} (\bibinfo {year} {2003})}\BibitemShut {NoStop}%
\bibitem [{\citenamefont {Beliaev}(1961)}]{Beliaev1961322}%
  \BibitemOpen
  \bibfield  {author} {\bibinfo {author} {\bibfnamefont {S.~T.}\ \bibnamefont
  {Beliaev}},\ }\href {\doibase 10.1016/0029-5582(61)90384-4} {\bibfield
  {journal} {\bibinfo  {journal} {Nucl. Phys.}\ }\textbf {\bibinfo {volume}
  {24}},\ \bibinfo {pages} {322 } (\bibinfo {year} {1961})}\BibitemShut
  {NoStop}%
\bibitem [{\citenamefont {Anguiano}\ \emph {et~al.}(2001)\citenamefont
  {Anguiano}, \citenamefont {Egido},\ and\ \citenamefont {Robledo}}]{Ang01}%
  \BibitemOpen
  \bibfield  {author} {\bibinfo {author} {\bibfnamefont {M.}~\bibnamefont
  {Anguiano}}, \bibinfo {author} {\bibfnamefont {J.~L.}\ \bibnamefont {Egido}},
  \ and\ \bibinfo {author} {\bibfnamefont {L.~M.}\ \bibnamefont {Robledo}},\
  }\href {\doibase 10.1016/S0375-9474(00)00445-0} {\bibfield  {journal}
  {\bibinfo  {journal} {Nucl. Phys. A}\ }\textbf {\bibinfo {volume} {683}},\
  \bibinfo {pages} {227 } (\bibinfo {year} {2001})}\BibitemShut {NoStop}%
\bibitem [{\citenamefont {Chabanat}\ \emph {et~al.}(1998)\citenamefont
  {Chabanat}, \citenamefont {Bonche}, \citenamefont {Haensel}, \citenamefont
  {Meyer},\ and\ \citenamefont {Schaeffer}}]{(Cha98)}%
  \BibitemOpen
  \bibfield  {author} {\bibinfo {author} {\bibfnamefont {E.}~\bibnamefont
  {Chabanat}}, \bibinfo {author} {\bibfnamefont {P.}~\bibnamefont {Bonche}},
  \bibinfo {author} {\bibfnamefont {P.}~\bibnamefont {Haensel}}, \bibinfo
  {author} {\bibfnamefont {J.}~\bibnamefont {Meyer}}, \ and\ \bibinfo {author}
  {\bibfnamefont {R.}~\bibnamefont {Schaeffer}},\ }\href@noop {} {\bibfield
  {journal} {\bibinfo  {journal} {Nucl. Phys. A}\ }\textbf {\bibinfo {volume}
  {635}},\ \bibinfo {pages} {231} (\bibinfo {year} {1998})}\BibitemShut
  {NoStop}%
\bibitem [{\citenamefont {Bartel}\ \emph {et~al.}(1982)\citenamefont {Bartel},
  \citenamefont {Quentin}, \citenamefont {Brack}, \citenamefont {Guet},\ and\
  \citenamefont {H{\aa}kansson}}]{(Bar82)}%
  \BibitemOpen
  \bibfield  {author} {\bibinfo {author} {\bibfnamefont {J.}~\bibnamefont
  {Bartel}}, \bibinfo {author} {\bibfnamefont {P.}~\bibnamefont {Quentin}},
  \bibinfo {author} {\bibfnamefont {M.}~\bibnamefont {Brack}}, \bibinfo
  {author} {\bibfnamefont {C.}~\bibnamefont {Guet}}, \ and\ \bibinfo {author}
  {\bibfnamefont {H.-B.}\ \bibnamefont {H{\aa}kansson}},\ }\href@noop {}
  {\bibfield  {journal} {\bibinfo  {journal} {Nucl. Phys. A}\ }\textbf
  {\bibinfo {volume} {386}},\ \bibinfo {pages} {79} (\bibinfo {year}
  {1982})}\BibitemShut {NoStop}%
\bibitem [{\citenamefont {Wang}\ \emph {et~al.}(2014)\citenamefont {Wang},
  \citenamefont {Dobaczewski}, \citenamefont {Kortelainen}, \citenamefont
  {Yu},\ and\ \citenamefont {Stoitsov}}]{PhysRevC.90.014312}%
  \BibitemOpen
  \bibfield  {author} {\bibinfo {author} {\bibfnamefont {X.~B.}\ \bibnamefont
  {Wang}}, \bibinfo {author} {\bibfnamefont {J.}~\bibnamefont {Dobaczewski}},
  \bibinfo {author} {\bibfnamefont {M.}~\bibnamefont {Kortelainen}}, \bibinfo
  {author} {\bibfnamefont {L.~F.}\ \bibnamefont {Yu}}, \ and\ \bibinfo {author}
  {\bibfnamefont {M.~V.}\ \bibnamefont {Stoitsov}},\ }\href {\doibase
  10.1103/PhysRevC.90.014312} {\bibfield  {journal} {\bibinfo  {journal} {Phys.
  Rev. C}\ }\textbf {\bibinfo {volume} {90}},\ \bibinfo {pages} {014312}
  (\bibinfo {year} {2014})}\BibitemShut {NoStop}%
\bibitem [{\citenamefont {Zhang}\ \emph {et~al.}(1989)\citenamefont {Zhang},
  \citenamefont {Casten},\ and\ \citenamefont {Brenner}}]{Zhang19891}%
  \BibitemOpen
  \bibfield  {author} {\bibinfo {author} {\bibfnamefont {J.-Y.}\ \bibnamefont
  {Zhang}}, \bibinfo {author} {\bibfnamefont {R.~F.}\ \bibnamefont {Casten}}, \
  and\ \bibinfo {author} {\bibfnamefont {D.~S.}\ \bibnamefont {Brenner}},\
  }\href {\doibase 10.1016/0370-2693(89)91273-2} {\bibfield  {journal}
  {\bibinfo  {journal} {Phys. Lett. B}\ }\textbf {\bibinfo {volume} {227}},\
  \bibinfo {pages} {1 } (\bibinfo {year} {1989})}\BibitemShut {NoStop}%
\bibitem [{\citenamefont {Cakirli}\ and\ \citenamefont
  {Casten}(2006)}]{Cakirli06}%
  \BibitemOpen
  \bibfield  {author} {\bibinfo {author} {\bibfnamefont {R.~B.}\ \bibnamefont
  {Cakirli}}\ and\ \bibinfo {author} {\bibfnamefont {R.~F.}\ \bibnamefont
  {Casten}},\ }\href {\doibase 10.1103/PhysRevLett.96.132501} {\bibfield
  {journal} {\bibinfo  {journal} {Phys. Rev. Lett.}\ }\textbf {\bibinfo
  {volume} {96}},\ \bibinfo {pages} {132501} (\bibinfo {year}
  {2006})}\BibitemShut {NoStop}%
\bibitem [{\citenamefont {Oktem}\ \emph {et~al.}(2006)\citenamefont {Oktem},
  \citenamefont {Cakirli}, \citenamefont {Casten}, \citenamefont {Casperson},\
  and\ \citenamefont {Brenner}}]{Cakirli06a}%
  \BibitemOpen
  \bibfield  {author} {\bibinfo {author} {\bibfnamefont {Y.}~\bibnamefont
  {Oktem}}, \bibinfo {author} {\bibfnamefont {R.~B.}\ \bibnamefont {Cakirli}},
  \bibinfo {author} {\bibfnamefont {R.~F.}\ \bibnamefont {Casten}}, \bibinfo
  {author} {\bibfnamefont {R.~J.}\ \bibnamefont {Casperson}}, \ and\ \bibinfo
  {author} {\bibfnamefont {D.~S.}\ \bibnamefont {Brenner}},\ }\href {\doibase
  10.1103/PhysRevC.74.027304} {\bibfield  {journal} {\bibinfo  {journal} {Phys.
  Rev. C}\ }\textbf {\bibinfo {volume} {74}},\ \bibinfo {pages} {027304}
  (\bibinfo {year} {2006})}\BibitemShut {NoStop}%
\bibitem [{\citenamefont {Qi}(2012)}]{Qi2012436}%
  \BibitemOpen
  \bibfield  {author} {\bibinfo {author} {\bibfnamefont {C.}~\bibnamefont
  {Qi}},\ }\href {\doibase 10.1016/j.physletb.2012.10.011} {\bibfield
  {journal} {\bibinfo  {journal} {Phys. Lett. B}\ }\textbf {\bibinfo {volume}
  {717}},\ \bibinfo {pages} {436 } (\bibinfo {year} {2012})}\BibitemShut
  {NoStop}%
\bibitem [{\citenamefont {Bonatsos}\ \emph {et~al.}(2013)\citenamefont
  {Bonatsos}, \citenamefont {Karampagia}, \citenamefont {Cakirli},
  \citenamefont {Casten}, \citenamefont {Blaum},\ and\ \citenamefont
  {Susam}}]{PhysRevC.88.054309}%
  \BibitemOpen
  \bibfield  {author} {\bibinfo {author} {\bibfnamefont {D.}~\bibnamefont
  {Bonatsos}}, \bibinfo {author} {\bibfnamefont {S.}~\bibnamefont
  {Karampagia}}, \bibinfo {author} {\bibfnamefont {R.~B.}\ \bibnamefont
  {Cakirli}}, \bibinfo {author} {\bibfnamefont {R.~F.}\ \bibnamefont {Casten}},
  \bibinfo {author} {\bibfnamefont {K.}~\bibnamefont {Blaum}}, \ and\ \bibinfo
  {author} {\bibfnamefont {L.~A.}\ \bibnamefont {Susam}},\ }\href {\doibase
  10.1103/PhysRevC.88.054309} {\bibfield  {journal} {\bibinfo  {journal} {Phys.
  Rev. C}\ }\textbf {\bibinfo {volume} {88}},\ \bibinfo {pages} {054309}
  (\bibinfo {year} {2013})}\BibitemShut {NoStop}%
\bibitem [{\citenamefont {Stoitsov}\ \emph {et~al.}(2007)\citenamefont
  {Stoitsov}, \citenamefont {Cakirli}, \citenamefont {Casten}, \citenamefont
  {Nazarewicz},\ and\ \citenamefont {Satu\l{}a}}]{PhysRevLett.98.132502}%
  \BibitemOpen
  \bibfield  {author} {\bibinfo {author} {\bibfnamefont {M.}~\bibnamefont
  {Stoitsov}}, \bibinfo {author} {\bibfnamefont {R.~B.}\ \bibnamefont
  {Cakirli}}, \bibinfo {author} {\bibfnamefont {R.~F.}\ \bibnamefont {Casten}},
  \bibinfo {author} {\bibfnamefont {W.}~\bibnamefont {Nazarewicz}}, \ and\
  \bibinfo {author} {\bibfnamefont {W.}~\bibnamefont {Satu\l{}a}},\ }\href
  {\doibase 10.1103/PhysRevLett.98.132502} {\bibfield  {journal} {\bibinfo
  {journal} {Phys. Rev. Lett.}\ }\textbf {\bibinfo {volume} {98}},\ \bibinfo
  {pages} {132502} (\bibinfo {year} {2007})}\BibitemShut {NoStop}%
\bibitem [{\citenamefont {Bender}\ and\ \citenamefont
  {Heenen}(2011)}]{PhysRevC.83.064319}%
  \BibitemOpen
  \bibfield  {author} {\bibinfo {author} {\bibfnamefont {M.}~\bibnamefont
  {Bender}}\ and\ \bibinfo {author} {\bibfnamefont {P.-H.}\ \bibnamefont
  {Heenen}},\ }\href {\doibase 10.1103/PhysRevC.83.064319} {\bibfield
  {journal} {\bibinfo  {journal} {Phys. Rev. C}\ }\textbf {\bibinfo {volume}
  {83}},\ \bibinfo {pages} {064319} (\bibinfo {year} {2011})}\BibitemShut
  {NoStop}%
\bibitem [{\citenamefont {Sato}\ \emph {et~al.}(2013)\citenamefont {Sato},
  \citenamefont {Dobaczewski}, \citenamefont {Nakatsukasa},\ and\ \citenamefont
  {Satu\l{}a}}]{Sato13}%
  \BibitemOpen
  \bibfield  {author} {\bibinfo {author} {\bibfnamefont {K.}~\bibnamefont
  {Sato}}, \bibinfo {author} {\bibfnamefont {J.}~\bibnamefont {Dobaczewski}},
  \bibinfo {author} {\bibfnamefont {T.}~\bibnamefont {Nakatsukasa}}, \ and\
  \bibinfo {author} {\bibfnamefont {W.}~\bibnamefont {Satu\l{}a}},\ }\href
  {\doibase 10.1103/PhysRevC.88.061301} {\bibfield  {journal} {\bibinfo
  {journal} {Phys. Rev. C}\ }\textbf {\bibinfo {volume} {88}},\ \bibinfo
  {pages} {061301} (\bibinfo {year} {2013})}\BibitemShut {NoStop}%
\bibitem [{\citenamefont {Sheikh}\ \emph {et~al.}(2014)\citenamefont {Sheikh},
  \citenamefont {Hinohara}, \citenamefont {Dobaczewski}, \citenamefont
  {Nakatsukasa}, \citenamefont {Nazarewicz},\ and\ \citenamefont
  {Sato}}]{PhysRevC.89.054317}%
  \BibitemOpen
  \bibfield  {author} {\bibinfo {author} {\bibfnamefont {J.~A.}\ \bibnamefont
  {Sheikh}}, \bibinfo {author} {\bibfnamefont {N.}~\bibnamefont {Hinohara}},
  \bibinfo {author} {\bibfnamefont {J.}~\bibnamefont {Dobaczewski}}, \bibinfo
  {author} {\bibfnamefont {T.}~\bibnamefont {Nakatsukasa}}, \bibinfo {author}
  {\bibfnamefont {W.}~\bibnamefont {Nazarewicz}}, \ and\ \bibinfo {author}
  {\bibfnamefont {K.}~\bibnamefont {Sato}},\ }\href {\doibase
  10.1103/PhysRevC.89.054317} {\bibfield  {journal} {\bibinfo  {journal} {Phys.
  Rev. C}\ }\textbf {\bibinfo {volume} {89}},\ \bibinfo {pages} {054317}
  (\bibinfo {year} {2014})}\BibitemShut {NoStop}%
\bibitem [{\citenamefont {Broglia}\ and\ \citenamefont
  {Zelevinsky}(2013)}]{(BroZel)}%
  \BibitemOpen
  \href@noop {} {\emph {\bibinfo {title} {Fifty Years of Nuclear BCS}}},\
  edited by
  \bibinfo {editor} {\bibfnamefont {R.~A.}\ \bibnamefont {Broglia}}\ and\
  \bibinfo {editor} {\bibfnamefont {V.}~\bibnamefont {Zelevinsky}}\
  (\bibinfo  {publisher} {World Scientific, Singapore},\ \bibinfo {year}
  {2013})\BibitemShut {NoStop}%
\end{thebibliography}

\end{document}